\documentclass[twocolumn,showpacs,preprintnumbers,amsmath,amssymb,pra,superscriptaddress]{revtex4}

\usepackage{graphicx}
\usepackage{dcolumn}
\usepackage{bm}

\def \eqnabbr{Eq.\ }

\begin{document}

\title{Dipole-field contributions to geometric-phase-induced false electric-dipole moment signals for particles in traps 
}
\author{P.G.\ Harris}\affiliation{Department of Physics and Astronomy, University of Sussex, Falmer, Brighton BN1 9QH, UK}
\author{J.M.\ Pendlebury}\affiliation{Department of Physics and Astronomy, University of Sussex, Falmer, Brighton BN1 9QH, UK}

\date{\today}

\begin{abstract}
It has been shown in an earlier publication \cite{pendlebury04} that magnetic field gradients applied to particles in traps can induce Larmor frequency shifts that may falsely be interpreted as electric-dipole moment (EDM) signals.  This study has now been extended to include nonuniform magnetic field gradients due to the presence of a local magnetic dipole.  It is found that, in the high orbit-frequency regime, the magnitude of the shifts can be enhanced beyond the simple expectation of proportionality to the volume-averaged magnetic-field gradient $\left<\partial B_z /\partial z\right>$.   

\end{abstract}

\maketitle

\section{Introduction}
Measurements of particle intrinsic electric-dipole moments (EDMs) are important because of the tight constraints that they impose upon theories that attempt to explain the origin of CP violation.  Such  measurements are generally made by applying to the particles of interest uniform static electric and magnetic fields that are, in turn, parallel and antiparallel to one another.  The Larmor precession frequency is measured, and any shift in frequency observed upon reversal of the electric field may ideally be attributed to an electric dipole moment.  However, when making such measurements,  great care must be taken to avoid systematic errors from a variety of sources.  One of the most important sources is the so-called  $\vec{v}\times\vec{E}$ effect, arising from the Lorentz transformation into the particle's rest frame of the laboratory-frame electric field:
\begin{equation}
\label{eqn:vxE}
\vec{B}_{\bot }^{\prime } =\gamma_L \left( \vec{B}-\frac{\vec{v}\times \vec{E}}{c^2}\right) _{\bot }.
\end{equation}

We consider here the case of particles stored in a trap with vertical ($z$) applied magnetic and electric fields $\vec{B}$ and $\vec{E}$.  The particles are moving slowly enough that the Lorentz factor $\gamma_L$ in \eqnabbr \ref{eqn:vxE} may be taken to be unity. The radial magnetic field components associated with a vertical gradient $\partial B_z/\partial z$ act in conjunction with the motion-induced field of \eqnabbr \ref{eqn:vxE} to produce a net rotating magnetic field of frequency $\omega_r$, which then shifts the Larmor frequency (initially $\omega_0$) of the trapped particles.  The induced shift is proportional to the applied electric field, and thus mimics the signal of an EDM.  Earlier publications \cite{pendlebury04, lamoreaux05} have analysed this geometric-phase effect in some detail.  

There are two separate situations to consider: first, the nearly adiabatic case of a slowly orbiting particle, $|\omega_r|<|\omega_0|$; and second, the non-adiabatic case of $|\omega_r|>|\omega_0|$.

\eqnabbr 34 of \cite{pendlebury04} shows that, for any shape of magnetic field, in the nearly adiabatic case the false EDM signal is 
\begin{equation}
\label{eqn:adiabatic_gp}
d_{af} = -\frac{J\hbar}{2}\left(\frac{\left<\partial B_{z}/\partial z\right>_V}{B_{z}^2}\right)\frac{v_{xy}^2}{c^2}\left[1-\frac{{\omega^*}_r^2}{\omega_0^2}\right]^{-1},
\end{equation}
where $v_{xy}$ is the velocity of the particle in the $xy$ (horizontal) plane, ${\omega^*}_r$ is the value of $\omega_r$ weighted appropriately to account for the populations of the various orbits, and the magnetic field gradient $\left<\partial B_{z}/\partial z\right>_V$ is averaged over the storage volume $V$.

The corresponding expression for the non-adiabatic case of $|\omega_r|>|\omega_0|$ (\eqnabbr 37 of \cite{pendlebury04}), for a {\em uniform} gradient $\partial B_z/\partial z$ and a cylindrical storage volume of radius $R$, is
\begin{equation}
\label{eqn:geophase_nonadiabatic}
d_{af} = \frac{J\hbar}{4}\left(\frac{\partial B_{z}}{\partial z}\right)\frac{\gamma^2R^2}{c^2}\left[1-\frac{{\omega}_0^2}{{\omega^\dagger}_r^2}\right]^{-1},
\end{equation}
where $\gamma$ is the gyromagnetic ratio and ${\omega^\dagger}_r$ is, as before, an appropriately weighted value of $\omega_r$.  It remained an open question in \cite{pendlebury04} as to whether the false EDM signal $d_{af}$ would always be proportional to the volume-averaged field gradient $\left<\partial B_{z}/\partial z\right>_V$ in the regime $|\omega_r|>|\omega_0|$.  In this study we show, by counter-example, that it is not.

\section{Geometric phase enhancement}

In the non-adiabatic case, the rate of addition of geometric phase is proportional to the radial component $B_r$ of the magnetic field.  For a uniform gradient this is given by
\begin{equation}
\label{eqn:Br_uniform}
B_r = - \frac{r}{2}\frac{\partial B_z}{\partial z}.
\end{equation}
The average $B_r$ over a circular region of radius $R$ in this case is therefore related to the gradient $ \partial B_{z}/\partial z $ by
\begin{equation}
\label{eqn:Br_unif_avg}
\frac{\left<B_r\right>}{\partial B_{z}/\partial z} = -\frac{R}{3}.
\end{equation}

We now consider the effect of a point magnetic dipole aligned along the $z$ axis.  The average radial magnetic field component $\left< B_r \right>$ over a circular region of radius $R$ at a height $z$ above the dipole is \footnote{\eqnabbr \ref{eqn:Br_dipole} is based on the standard dipole field at distances much larger than the scale of the dipole itself, and cannot be used in regions too close to the dipole.  For any dipole, by symmetry, $ B_r = 0$ everywhere in the plane $z=0$.}
\begin{equation}
\label{eqn:Br_dipole}
\left<B_r\right> = \left( \frac{p}{4\pi}\right)\frac{2R}{z(z^2+R^2)^{3/2}},
\end{equation}
where $p$ is the dipole moment.  The average gradient $\left< \partial B_{z}/\partial z \right>$ over the same region is
\begin{equation}
\label{eqn:avg_grad_dipole}
\left< \partial B_{z}/\partial z \right> = 
\left( \frac{p}{4\pi}\right)\frac{-6z}{(z^2+R^2)^{5/2}}.
\end{equation}
It follows that $\left<B_r\right>$ is now related to $\left< \partial B_{z}/\partial z \right>$ by
\begin{equation}
\label{eqn:Br_grad_dipole}
\frac{\left<B_r\right>}{\left< \partial B_{z}/\partial z \right>} = 
-\frac{R}{3}\left(1+\frac{R^2}{z^2}\right).
\end{equation}
This is clearly enhanced by a factor $(1+R^2/z^2)$ beyond the relation of \eqnabbr \ref{eqn:Br_unif_avg}, 
and the expression (\ref{eqn:geophase_nonadiabatic}) above for the geometric-phase-induced false EDM signal would therefore be expected to be enhanced by the same factor, with appropriate averaging over the range of $z$.

This effect has been studied with a numerical simulation, using a storage volume of height 12 cm and radius 23.5 cm (corresponding to the configuration of the neutron EDM experiment at the Institut Laue-Langevin, Grenoble).  The results for several values of the distance $z_1$ of the dipole below the floor of the storage volume are shown in Table \ref{tab:sim_results}. 
\begin{table}[bp] 
\begin{tabular}{|l|l|l|l|}
\hline
$z_1$ (cm) & Enhancement & Expected & Ratio\\
      & (simulation)& enhancement   &      \\ 
\hline
5  & 4.25 & 7.50 & 0.57 \\ \hline
10 & 2.56 & 3.51 & 0.73 \\ \hline
20 & 1.76 & 1.86 & 0.95 \\ \hline
30 & 1.42 & 1.44 & 0.99 \\ \hline
\end{tabular}
\caption{Comparison of the geometric-phase enhancement factors due to a point magnetic dipole at various distances below the floor of a cylindrical storage volume of height 12 cm and radius 23.5 cm, obtained from a numerical simulation, with the factors expected from \eqnabbr \ref{eqn:Br_grad_dipole}}
\label{tab:sim_results}

\end{table}%

Up to an enhancement factor of about two, the the simulation and the $z$-averaged prediction of \eqnabbr \ref{eqn:Br_grad_dipole} agree to within a few percent.  However, when the dipole is in close proximity, the enhancement factor is seen to be suppressed relative to the analytic prediction.  We do not have a quantitative explanation of the underlying mechanism for this reduction; however, we believe that it may be due to the increasing rapidity of the changes of magnitude of the radial magnetic field component.  As the dipole approaches the storage volume, $B_r$ becomes much more localized compared to the size of the radius $R$ of the cell. Under these conditions the moving particles see more of the amplitude $B_r$ being assigned to higher harmonics of the particle orbit frequency.  As these are further away from the (relatively low) Larmor frequency $\omega_0$, they are less effective in producing a frequency shift, and the enhancement is thereby suppressed. 

The adiabatic case was also simulated with the same magnetic-dipole field environment.   No departure from \eqnabbr \ref{eqn:adiabatic_gp} was observed.

\section{Conclusion}

At current and future anticipated levels of sensitivity, geometric-phase-induced false EDM signals are an important potential source of systematic errors in EDM measurements.  It has been shown in this Report that non-uniform magnetic field gradients can, in the non-adiabatic case $|\omega_r|>|\omega_0|$, result in an enhancement of the false EDM signal above and beyond that anticipated by a simple dependence upon the volume-averaged gradient.  A simple formula has been derived for the situation of a dipole aligned with the symmetry axis of the storage volume; however, this formula overestimates the enhancement when the dipole is in close proximity to the bottle, and at the current level of understanding it appears to be necessary to rely upon numerical simulations to obtain accurate predictions of this effect.

\acknowledgments
The authors are most grateful to their colleagues on the neutron EDM experiment for valuable discussions. This work was supported in part by grant no.\ PP/B500615/1 from the UK Particle Physics and Astronomy Research Council.  

\bibliography{neutron_edm}

\end{document}